\documentclass[%
 reprint,
 amsmath,amssymb,
 aps,
prb,
]{revtex4-2}

\usepackage{graphicx}% Include figure files
\usepackage{dcolumn}% Align table columns on decimal point
\usepackage{bm}% bold math
\begin{document}

%\preprint{APS/123-QED}

\title{Phases of $^4$He and  H$_2$ adsorbed on a single carbon nanotube}

\author{M. C. Gordillo$^{1,2}$ and R. Rodr\'{\i}guez-Garc\'{\i}a$^1$}
\affiliation{$^{1}$Departamento de Sistemas F\'{\i}sicos, Qu\'{\i}micos
y Naturales, Universidad Pablo de
Olavide, Carretera de Utrera km 1, E-41013 Sevilla, Spain}
\affiliation{$^2$Instituto Carlos I de Física Teórica y Computacional, Universidad de Granada, E-18071 Granada, Spain}

\author{J. Boronat$^3$}
\affiliation{$^3$Departament de F\'{\i}sica,
Universitat Polit\`ecnica de Catalunya,
Campus Nord B4-B5, 08034 Barcelona, Spain}

\date{\today}% It is always \today, today,

\begin{abstract}
Using a diffusion Monte Carlo (DMC) technique, we calculated the phase diagrams of $^4$He and H$_2$ adsorbed 
on a single (5,5) carbon nanotube, one of the narrowest that can be obtained experimentally.  For a single monolayer,  when 
the adsorbate density increases,  both species undergo a series of first order 
solid-solid phase transitions between incommensurate arrangements. Remarkably,  
the  $^4$He lowest-density  solid phase shows supersolid behavior in contrast 
with the normal solid that we found for H$_2$.   The nature of the second-layer 
is also different for both adsorbates.  Contrarily to what happens on graphite, 
 the second-layer of  $^4$He on that tube is a liquid, at least up to the 
density corresponding to a third-layer promotion on a flat substrate. However, 
the second-layer of H$_2$ is a solid that, at  its 
lowest stable density, has a small but observable superfluid fraction.   
 \end{abstract}

\maketitle
               
\section{Introduction}  
A carbon nanotube can be thought of as the result of folding a single graphene 
sheet over itself to form a seamless cylinder~\cite{iijima}.  It is then a 
quasi-one dimensional (1D) structure that can be coated with different kinds of 
adsorbates.  
In principle, that reduced dimensionality could produce phase 
diagrams different than those on quasi-two dimensional (2D) environments such as 
graphite or graphene.  The study of the adsorption capabilities of those 
carbon cylinders is possible since 
an isolated carbon nanotube can be synthesized and made to work as 
a mechanical resonator. Its resonant frequencies  change  
upon loading, allowing in this way an accurate determination of the adsorbate 
phases~\cite{cobden1,cobden2,chiu,chaste1,tavernarakis,noury, todoshchenko}.  
In principle, this resonant frequency can be monitored to check if we have 
a supersolid structure, as it was done for the second layer of $^4$He on graphite
\cite{nyeki,kim}. 

Experimental studies on Ne,  Ar, and Kr~\cite{cobden1,cobden2,tavernarakis} 
indicate that the first layers of those gases on nanotubes are qualitatively 
similar to those found in their flat equivalents, the only difference being the 
smaller binding energies on the cylinders due to their curved nature 
\cite{cobden1,cobden2}.  On the other hand, it is known that $^4$He on a 
single nanotube is adsorbed on a layer-by-layer process,  similar to the 
deposition on graphite~\cite{noury}.  The main goal of our present work is 
to deeply study the behavior of $^4$He and H$_2$ on top of a (5,5) carbon 
nanotube, both  the first and second layers.  We chose that tube since its 
narrowness, with a radius of $3.42$ \AA,  makes any difference with a flat 
graphite (or graphene) substrate larger than for thicker tubes.  In 
particular,  we were interested to study if the imposed curvature produced any 
additional phase, such as the supersolids 
experimentally detected \cite{nyeki,fukuyama,kim} and theoretically predicted in the first 
\cite{claudiosuper} and second layer of $^4$He on a carbon substrate 
\cite{prl2020}, or in the second layer of H$_2$ on graphite \cite{prb2022}.  
That superfluidity, in the first layer solids,  was not calculated  
in previous studies \cite{tubo2,tubo1}.  We also explored the possibility 
of second-layer liquid phases for both species.

The rest of the paper is organized as follows. The diffusion Monte Carlo method 
used in our study is discussed in Sec. II. Sec. III comprises the results 
obtained, with special attention to the stable phases of both $^4$He and H$_2$ 
adsorbed on the nanotube. The results for the superfluid fraction were 
calculated by the standard winding number estimator, in the limit of zero 
temperature. Finally, the main conclusions and discussion of the results are 
contained in Sec. IV.

\section{Method}

To study the stability of the different phases, we calculated the respective 
ground states ($T=0$) of  $^4$He atoms/H$_2$ molecules on a corrugated 
carbon nanotube at several densities.  This means to write down and solve the 
many-body Schr\"odinger equation 
that describes the adsorbate. Following previous works 
in similar systems,  the Hamiltonian could be written down as
\begin{equation} \label{hamiltonian}
  H = \sum_{i=1}^N  \left[ -\frac{\hbar^2}{2m} \nabla_i^2 +
V_{{\rm ext}}(x_i,y_i,z_i) \right] + \sum_{i<j}^N V_{{\rm pair}}
(r_{ij}) \ ,
\end{equation}
where $x_i$, $y_i$, and $z_i$ are the coordinates of the each of the $N$ adsorbate particles with mass $m$.
$V_{{\rm ext}}(x_i,y_i,z_i)$ is the interaction potential between each atom or molecule and 
all the individual carbon atoms in the nanotube, that is considered to be a 
rigid structure.  Those potentials are of  
Lennard-Jones type, with standard parameters taken from Ref. 
\onlinecite{carlosandcole} in the case of  $^4$He-C,  and from 
Ref. \cite{coleh2} for the H$_2$-C interaction.  
$V_{\rm{pair}}$ accounts for the $^4$He-$^4$He and H$_2$-H$_2$ interactions (Aziz \cite{aziz} and  
Silvera and Goldman \cite{silvera} potentials, respectively), that depend 
only on the distance $r_{ij}$ between particles $i$ and $j$. Both potentials are 
the standard models in previous literature.  

To actually solve the many-body Schr\"{o}dinger equation derived from
the Hamiltonian in Eq. \ref{hamiltonian},  we resorted to the diffusion Monte 
Carlo (DMC) algorithm.
This stochastic numerical technique obtains, within some statistical 
noise, relevant ground-state properties  of the $N$-particle system. To 
guide the diffusion 
process involved in DMC, one introduces a trial wave function which acts as 
importance sampling, reducing the variance to a manageable level. 
%However, the 
%DMC algorithm does not merely sample the configurations corresponding to the 
%square
%of the trial function, but it is able to correct the energies to obtain the 
%real  
%ground state of the system \cite{borobook}.   
%With the help 
%of a proper trial wave function, we can fix the particular phase that we are 
%going to consider. 
The trial wave functions used in our work derive from similar forms used 
previously in DMC calculations of $^4$He and H$_2$ on graphene and graphite 
\cite{prl2009,prb2010,prl2020,prb2022},
that were able to reproduce available experimental data 
\cite{crowell1,crowell2,greywall1,greywall2,nyeki,kim,wiechert}. 
We are then confident that the trial wave functions used in the present 
work will also provide a reasonable description of the adsorbates on a similar, 
albeit curved, substrate.
    
In the present case, the trial wave function was built as a product of two 
terms. The first one is of Jastrow  type between the adsorbate particles,
\begin{equation}
\Phi_J({\bf r}_1,\ldots,{\bf r}_N) = \prod_{i<j}^{N} \exp \left[-\frac{1}{2}
\left(\frac{b}{r_{ij}} \right)^5 \right] \,
\label{sverlet}
\end{equation}
with $b$ a variationally optimized parameter, whose
values were $3.07$ \AA$ $ for the $^4$He-$^4$He case \cite{tubo2} and $3.195$ 
\AA$ $ for the H$_2$-H$_2$ pair \cite{tubo1}.
The second part incorporates the presence of the C atoms and localization 
terms, 
\begin{eqnarray}
\Phi_s({\bf r}_{1},\ldots,{\bf r}_{N})  = 
\prod_i^{N}  \prod_J^{N_C} \exp \left[ -\frac{1}{2} \left( \frac{b_{{\text
C}}}{r_{iJ}} \right)^5 \right] \nonumber \\
 \times \prod_{I=1}^{N} \left[ \sum_{i=1}^{N} \exp
\{-c_{1,2} [({\bf r}_i-{\bf r}_{\text{site},I})^2] \} \ \right] 
\times \prod_i^{N} \Psi(R_i).  \ \ \ \ \   \ 
\label{t2}
\end{eqnarray}
Here, $N_C$ stands for the number of carbon atoms of the nanotube while the 
parameters $b_C$ were taken from previous calculations on the same substrates 
\cite{tubo2,tubo1}.   $r_{iJ}$ are the distances between a particle $i$ ($^4$He 
or H$_2$), and a carbon atom, $J$.  $\Psi(R_i)$ is a one-body function that 
depends on the radial distance of those particles to the central axis of the 
tube, $R_i$.  For an atom or molecule located on the layer closest to the carbon 
substrate, those were taken from Refs. \onlinecite{tubo2} and 
\onlinecite{tubo1} for $^4$He and H$_2$. Those functions have maxima located at distances from the center of 
6.26 and 6.36 \AA, respectively. On the other hand, and following the procedure already used for graphite
\cite{prl2020,prb2022,prbsecondh2},  we radially confined particles in 
the second layer  by Gaussian functions, with  variationally-optimized   
parameters.   

The remaining part of Eq. \ref{t2} allows us to distinguish between a 
translationally invariant liquid ($c_{1,2}$= 0) and a (super)solid phase 
($c_{1,2} \ne$ 0). Labels 1 and 2 stand for  first and second layers,  
respectively.   For solids,  the values of $c_{1,2}$ were variationally 
optimized to obtain the minimum energies for each species and density.  
For particles in the first layer, the parameters for the solid where found to be identical 
to those obtained for flat graphene and used in previous calculations for similar
systems \cite{tubo2,tubo1}. This means we used  linear extrapolations between $c_1$ = 0.15\AA$^{-2}$ (for a density of 0.08 \AA$^{-2}$) and
$c_1$ = 0.77\AA$^{-2}$ (for 0.1 \AA$^{-2}$) for $^4$He \cite{prl2009} and 
between $c_1$ = 0.61\AA$^{-2}$ (for 0.08 \AA$^{-2}$) and
$c_1$ = 1.38\AA$^{-2}$  (for 0.1 \AA$^{-2}$) in the case of H$_2$ \cite{prb2010}. For lower densities we used the
value corresponding to 0.08\AA$^{-2}$ and for larger ones, the one corresponding 
to 0.1 \AA$^{-2}$. For the second layer of H$_2$, $c_2$ = 0.46 \AA$^{-2}$ was used for all densities.     
The optimal $c_2$ parameter $^4$He was $c_2$=0, that corresponds to a liquid (see below).
For 
any value of $c_{1,2}$, the form of the trial function allows the $^4$He atoms 
and the H$_2$ molecules 
to be involved in exchanges and recover indistinguishability,  a necessary ingredient to model a supersolid \cite{claudiosuper}. 
Alternatively,  for solid phases one can use a simplified version of Eq. \ref{t2} in which each particle is pinned 
to a single crystallographic site. This is the ansatz used to describe first-layer phases in Refs. \cite{tubo2,tubo1} and would 
produce, by construction, normal solids. When we used that approximation, we obtained higher or equal energies per particle
than when we use Eq. \ref{t2}. The case of the equal values for the energies correspond to cases in which the superfluid 
estimator is equal to zero, i.e., when we recover the normal behavior of the solid.      
%The values of $c_1$ and $c_2$ are different.  In the most extreme case, we can have a liquid 
%on top, i.e. , $c_2 = 0$, and a solid under it with $c_1 \neq 0$. 

Finally, in Eq. \ref{t2}, $x_{\text{site}},y_{\text{site}}$, and 
$z_{\text{site}}$ are the 
crystallographic positions that define the solids we wrap around the (5,5) tube. 
 Those are  incommensurate arrangements
built up by locating a given number, $n$, of adsorbate particles on planes perpendicular to the main axis of the 
nanotube.  
%In between a pair of  those particle "bracelets'' we introduced 
%another one with the positions of the atoms/molecules rotated to be in the 
%middle of two particles in neighboring rows.  The density of the solid is 
%changed by varying the distance between parallel rows of atoms. The phases so 
%constructed are labeled as  $n$-in-a-row solids, both in the first and second 
%layers. Alternatively, 
One can visualize those phases by imagining that the coated 
cylinder is fully cut longitudinally to have a long rectangle in which the 
shorter side 
corresponds to the length of the circumference that defines the tube. In the case 
of a first $^4$He layer, this means 2$\pi \times$ 6.26 $ $\AA (see above). 
On that short side, we locate $n$ atoms or molecules uniformly spaced, with a distance 
between them in the transverse (short) direction of  $d_t$\AA, and 
added as many parallel rows as the length of the tube will allow. Those arrays of
atoms will be separated by a longitudinal distance of $d_l$ \AA. 
Between those rows, we will include another one separated 
$d_l$/2 from the contiguous lines, and whose particles will be displaced $d_t$/2 
in transverse direction with respect to the previous and following lines.  
A similar 
procedure is used to build the solids in the second layer, located at an averaged 
distances for the center or the tube of 8.98 and 9.42 \AA$ $ for $^4$He and H$_2$, 
respectively. Since we started 
by $n$ particles located on rows in the transverse direction, following Refs. 
\onlinecite{tubo1} and \onlinecite{tubo2}, we name these phases $n$-in-a-row 
solids.   A picture of one of those phases can be found in Ref. 
\onlinecite{tubo1} for the case of H$_2$.

The data presented in this paper are the mean  of ten independent DMC 
simulations, and
the error bars, when shown, correspond to the variance of these calculations. 
Every DMC history consists of 1.2$\times$10$^5$ steps
involving a movement of all the particles of each of the 300 replicas 
(walkers) that describe the different configurations. Larger number of walkers or 
longer simulations do not change the averages given. The values of the observables presented 
here were calculated after equilibration  (2$\times$10$^4$ time steps), i.e., 
when no obvious drift in their values 
as a function of the simulation time was detected.  We made the simulations on 
isolated 
tubes, what means that periodic boundary conditions were applied only on the 
direction 
corresponding to the length of the tube. The number of particles in the simulation cells 
and their lengths were varied to produce the desired densities for any of the two adsorbates.
The first number was between a minimum of 100 (for a one-layered tube) to a maximum of 
364 for the highest densities considered in this work. However, some simulations 
including up to 500 particles were made to verify that our results were 
not affected by size effects. Conversely, the lengths of the simulation cells 
varied in the range 33-45 \AA.

\section{Results}

\subsection{$^{\bf 4}$He}

\begin{figure}
\begin{center}
\includegraphics[width=0.8\linewidth]{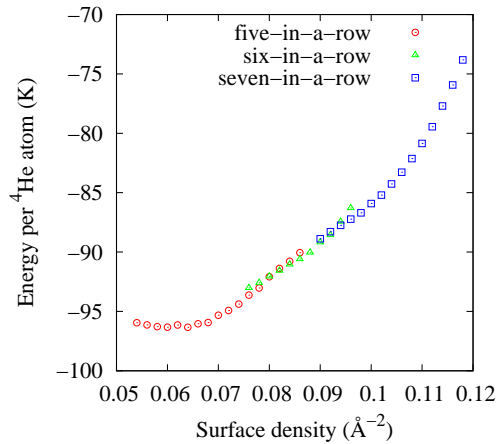}
\caption{Energy per $^4$He atom as a function of the density, for $n$-in-a-row 
($n= 5,6,7$) first layer solids.} 
\label{enerhe4bajo}
\end{center}
\end{figure}

We first study the phase diagram of $^4$He on a (5,5) 
nanotube, including the promotion from a first to a second 
layer.  Fig. \ref{enerhe4bajo} gives us the energy per particle for the 
different first-layer incommensurate solids.  Those extend the results given in 
Ref. \onlinecite{tubo2} in a double way.  First, we consider now larger Helium 
densities in the 7-in-a-row solid, what would allow us to study all the possible 
first-order transitions between incommensurate arrangements up to the 
second-layer promotion.  
In addition, since those first-layer solids are described by Eq. \ref{t2}, 
instead of having each particle of the solid pinned to a single 
crystallographic position (Nosanow-Jastrow wave function),  we can  access 
to any possible supersolid phases.  

The energy per particle for any of the three $n$-in-a-row $^4$He 
($n=5,6,7$) solids are shown in Fig.~\ref{enerhe4bajo}.  
As indicated above,  to calculate the surface densities we used cylinders with a radius given by the average distance of the $^4$He atoms to the
center of the tube, in this case $6.26$ \AA. We do not show in 
Fig.~\ref{enerhe4bajo}   the results of a translationally invariant liquid phase 
because the energies for that phase are above the ones shown in that figure.  
The same applies to the  $2/5$,  $3/7$, and $\sqrt{3} \times  \sqrt{3}$ 
registered solids \cite{greywall2}.  By means of a least-squares fit method to 
the 5-in-a-row data, we can obtain the
minimum density and its corresponding energy per particle. The results are given
in Table \ref{tab:table1}.  There, we can see that they are slightly different 
from those of Ref.~\onlinecite{tubo2},  with $\rho = 0.062$ \AA$^{-2}$ and $E = 
-96.10$ K  The reason for the discrepancy is the 5-in-a-row solid being a 
supersolid instead of the normal solid previously considered.  Its superfluid 
density was calculated using the zero-temperature winding number estimator, 
derived in Ref. \onlinecite{gubernatis},
\begin{equation} \label{super}
\frac{\rho_s}{\rho}= \lim_{\tau \to \infty} \alpha \left(
\frac{D_s(\tau)}{\tau} \right) \ ,
\end{equation}
with $\tau$ the imaginary time used in the quantum Monte Carlo
simulation. Here, $\alpha = N/(2 D_0)$, with $D_0 = \hbar^2/(2m)$, and
$D_s(\tau) = \langle [{\bf R}_{CM}(\tau)-{\bf R}_{CM}(0)]^2 \rangle$. ${\bf
R}_{CM}$ is the
position of the center of mass of the $N$ $^4$He atoms in the simulation 
box, using only their $z$ coordinates  on which periodic boundary conditions 
are applied. 
%This estimator is different than the one defined by Ceperley and
%Pollock for finite temperature systems \cite{ceperleypollock}. In any case, 
%for finite temperature, one would expect that last estimator to be zero, 
%as befits a (quasi)-1D system. That limit does not hold for T=0.

\begin{table}[b]%The best place to locate the table environment is directly after its first reference in text
\caption{\label{tab:table1}
Lowest and highest stability density limits for the different single-layer solid phases adsorbed on a (5,5) carbon nanotube.  
We include also the lowest stable density limit for the second layer of $^4$He,  a liquid,  and of H$_2$, a solid.  The error bars are given in parenthesis and correspond to the last decimal place given. } 
\begin{ruledtabular}
\begin{tabular}{cccc}
$\rho$ ($^4$He)  (\AA$^{-2}$) & E ($^4$He) (K) & $\rho$ (H$_2$) (\AA$^{-2}$) & E (H$_2$) (K) \\
%\textrm{Right}\\
\colrule
5-in-a row   \\
\colrule 
0.0605(5)  &  -96.5(1) & 0.062(2)  & -349.0(1) \\
0.076(1) & -93.7(1)  & 0.068(2)  & -346.1(1)  \\
\colrule
6-in-a-row \\
\colrule
0.086(2)  &  -90.6(1) & 0.080(2)  & -334.1(1) \\
0.088(1) & -90.0(1)  & 0.085(2)  & -326.7(1)  \\
\colrule
7-in-a-row \\
\colrule
0.096(2)  &  -87.2(1) & 0.0925(5)  & -314.3(1) \\
0.110(2)  &  -80.9(1) & 0.0975(5)  & -304.1(1) \\
\colrule
second layer \\
\colrule
0.181(1)  &  -48.9(1) & 0.166(2)  & -199.7(1) \\
\end{tabular}
\end{ruledtabular}
\end{table}

\begin{figure}
\begin{center}
\includegraphics[width=0.8\linewidth]{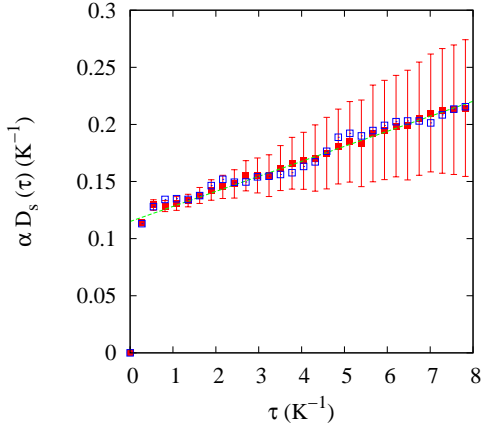}
\caption{Estimator of the superfluid density for the first-layer 5-in-a-row supersolid $^4$He phase at $\rho$ = 0.062 \AA$^{-2}$ (solid squares) and $\rho$ = 0.062 \AA$^{-2}$ (open squares). The straight line is a 
linear least-squares fit to the $\rho$ = 0.062 \AA$^{-2}$  data for 
$\tau >$  3 K$^{-1}$. }
\label{supersolidohe4}
\end{center}
\end{figure}

The results of the superfluid estimator to the 5-in-a-row solid can be 
seen in Fig \ref{supersolidohe4} for 
$\rho = 0.062$ \AA$^{-2}$ (solid squares) and 0.064 \AA$^{-2}$ (open squares). 
The line is a least-square fitting 
to $\alpha D_s (\tau)$ versus $\tau$  for the lowest density in the range 3 $< 
\tau <$ 8 K$^{-1}$. 
The value obtained is $\rho_s/\rho$ = 1.31$\pm$0.05 \% for  0.062 \AA$^{-2}$  and  $\rho_s/\rho$ = 1.35$\pm$0.05 \% 
(line not shown) for  0.064 \AA$^{-2}$.  In both cases, the superfluid fraction 
is of the same order, but slightly larger, that 
the one found for a registered phase of $\rho$ = 0.0636 \AA$^{-2}$ of $^4$He on graphene
($\rho_s/\rho$ = 0.67$\pm$0.01 \%) \cite{claudiosuper}. No significant size 
effects were found for larger simulation
cells, in accordance with results in Ref. \onlinecite{prb2022}.

Using the data reported in Fig.~\ref{enerhe4bajo}, and by means of 
double-tangent Maxwell constructions, we can obtain the stability limits 
of the different first-layer $^4$He solid phases. The results are given in Table \ref{tab:table1}, where we can find 
the lowest and highest stable densities, $\rho$,  for those arrangements.  This means that, for instance,  there is a first-order 
phase transition between a 5-in-a-row solid of $\rho$ = 0.076 \AA$^{-2}$ and a 6-in-a-row one with $\rho$ = 0.086 \AA$^{-2}$
and the same can be said for a  6-in-a-row of $\rho$ = 0.088 \AA$^{-2}$ and 7-in-a-row $\rho$ = 0.096 \AA$^{-2}$
structures.  Both 6- and 7-in a row solids are normal solids,  with  $\rho_s/\rho$ = 0. 

If we keep increasing the $^4$He density on top of the nanotube,  
it is promoted to a second layer.  To 
obtain the stability range of that second layer,  we used the same technique as 
in Ref.~\onlinecite{prbsecondhe4}, i.e., we calculated the energies per particle 
for arrangements in which the first layers were 7-in-a-row solids with different 
densities and put 
on top an increasing number of $^4$He atoms.  Then,  the stable second-layer 
will be the one with the lowest energy for a given
total density.  That density is calculated as the sum of the first-layer one, in 
which we used as an adsorbent surface a cylinder or radius 6.26 \AA, and the one 
corresponding to the second layer, computed using another cylinder whose 
distance from the axis of the tube was 8.98 \AA. 
%This is the average distance of 
%a set of $^4$He atoms on the second layer.   
Both radii corresponds to the mean 
distances of the atoms to the center of the tubes.  

\begin{figure}
\begin{center}
\includegraphics[width=0.8\linewidth]{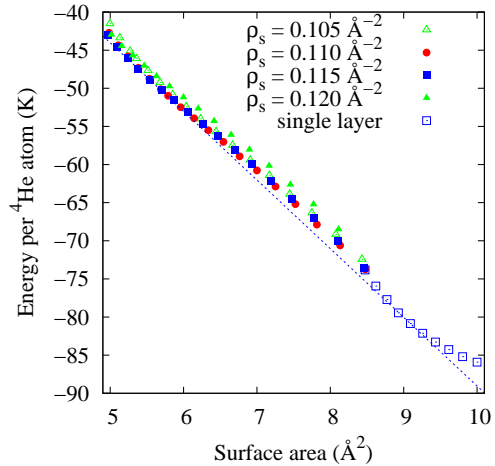}
\caption{Energy per $^4$He atoms as a function of the inverse of the total $^4$He density in the range in which we can have first-only and first and second adsorbate layers.  Full squares, 7-in-a row solids of different densities.  Solid symbols, two-layer structures with different values of the densities 7-in-a-row solids (shown) underneath. In all cases, the second layers were translationally invariant liquid structures. The dotted line is a Maxwell construction between the stable densities of first-only and first+second layers. The error bars are of the size of the symbols and not shown for simplicity. }
\label{enerhe4liq}
\end{center}
\end{figure}

The results of the 
calculations for different $^4$He loadings are displayed in 
Fig.~\ref{enerhe4liq}.  
In that figure,  we can see that the lowest two-layer structure has a  
7-in-a-row first layer solid underneath with $\rho$= 0.110 \AA$^{-2}$ (full 
circles) for low densities, and changes to a similar structure with $\rho$= 
0.115 \AA$^{-2}$ upon loading (full squares).  
In the $x$-axis we show the inverse of the density, or  
surface area, since in that way we can show the
double-tangent Maxwell construction between a single layer solid and the liquid on top of it.  That construction,  shown in 
Fig. \ref{enerhe4liq} as a dotted line, indicates that the coexistence region is between total densities of 0.110 \AA$^{-2}$ and 0.181 \AA$^{-2}$ (see Table \ref{tab:table1}).  For that last structure, the density of the solid close to the nanotube is $\rho$= 0.115 \AA$^{-2}$. 
This means that, as in the case of the $^4$He adsorption on graphite, there is a compression of the layer close to the carbon when a second adsorbate sheet is deposited on top of it \cite{prbsecondhe4}.  

\begin{figure}
\begin{center}
\includegraphics[width=0.8\linewidth]{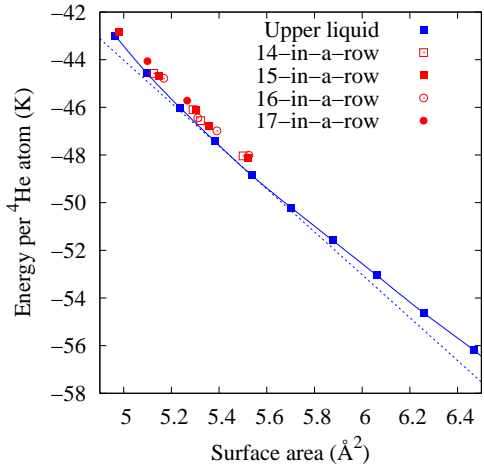}
\caption{Same as in Fig.~\ref{enerhe4liq}, but for surface areas in the range 
5-6.5 \AA$^2$.  The dotted line is also the same Maxwell construction between 
the stable densities of first-only and first+second layers.  The full line is a 
guide-to-the-eye to the results of a structure in which the second layer is a 
liquid.  We also show simulation results for different unstable $n$-in-a-row 
second layer-solids.  The error bars are of the size of the symbols. }
\label{enerhe4liqsol}
\end{center}
\end{figure}

To see more clearly the lowest stability limit of the two-layer structure, we display in Fig. \ref{enerhe4liqsol} a blown-up version of Fig. \ref{enerhe4liq} for total surface areas in the range 5-6.5 \AA$^2$.  The first value corresponds to the inverse of the experimental density for the promotion to a third layer of $^4$He on graphite \cite{nyeki}.  It displays the same Maxwell construction line as in the previous figure as a dotted line.   We included also the results for several $n$-in-a row solid structures, with $n$ in the range 14-17.  Those data were obtained  fixing the $c_2$ parameter in Eq. \ref{t2} to 0.30 \AA$^{-2}$, but they are qualitatively similar to any simulation results in which 
$c_2 \neq$ 0, i.e., arrangements with two solid layers are always metastable 
with respect to other with a liquid surface on top.  In this, the behavior of 
$^4$He is different than of its flat counterpart adsorbed on graphite, that 
solidifies before a third-layer promotion 
\cite{greywall1,greywall2,crowell1,crowell2}. Obviously, that second-layer 
liquid is a superfluid, since the value of $\rho_s/\rho$ obtained applying Eq. 
\ref{super} is 1. 
 
\subsection{H$_2$}

We will turn now our attention to H$_2$,  dealing first with first-layer 
solids described by Eq. \ref{t2} that does not pin 
the H$_2$ molecules to a single crystallographic position, as it was done in 
Ref.  \onlinecite{tubo1}.  This open the possibility of having supersolids,  
possibility that, as we will see below, is not fulfilled. 

\begin{figure}
\begin{center}
\includegraphics[width=0.8\linewidth]{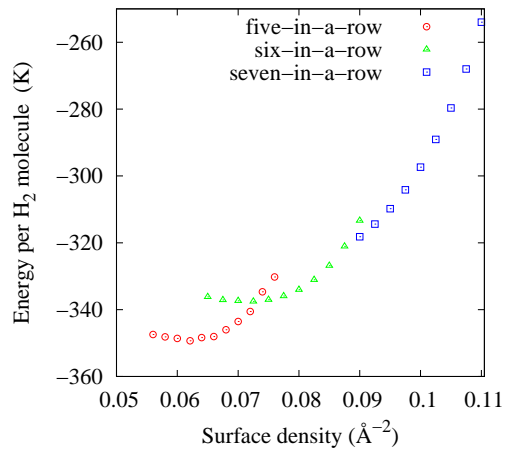}
\caption{Same as in Fig. \ref{enerhe4bajo} for first-layer H$_2$ solids. }
\label{enerh2bajo}
\end{center}
\end{figure}

\begin{figure}
\begin{center}
\includegraphics[width=0.8\linewidth]{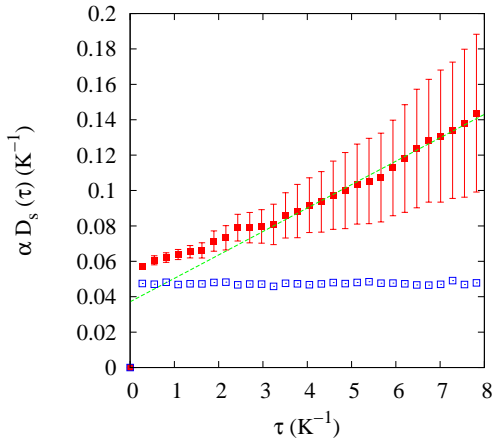}
\caption{Estimator of the superfluid density for the first-layer 5-in-a-row supersolid H$_2$ phase at $\rho$ = 0.062 \AA$^{-2}$ (open squares) and for a second-layer solid at total density $\rho$ = 0.167 \AA$^{-2}$ (solid squares).  The straight line is a 
linear least-squares fit to the latter set of data for the range
$\tau >$  3 K$^{-1}$. }
\label{supersolidoh2}
\end{center}
\end{figure}

Fig. \ref{enerh2bajo} is the H$_2$ counterpart of the Fig. \ref{enerhe4bajo} 
for $^4$He. The densities in the $x$-axis are calculated now using 
adsorbate cylinders of radius 6.36 \AA. This difference is due to the larger
size of the H$_2$ molecule with respect to the $^4$He atom.  Making use of the 
double tangent Maxwell construction lines between those solids, we find a range 
of densities, shown in Table \ref{tab:table1},  for which we have stable 
5-in-a-row and 6-in-a-row solids.  In this case, and 
contrarily to  what happens to $^4$He, both the stability limits and the 
energies per particle are basically identical to those found in Ref. 
\onlinecite{tubo1}.  The reason can be understood by looking at Fig. 
\ref{supersolidoh2}.  There, we can see (open symbols) the same estimator for 
the superfluid density shown in Fig. \ref{supersolidohe4} but  for a first-layer 
H$_2$ solid at $\rho$ = 0.062 \AA$^{-2}$, the minimum of the 5-in-a-row energy 
curve in Fig.  \ref{enerh2bajo}.  We can see that  $\alpha$ D$_s 
(\tau)$ versus $\tau$  is constant for large $\tau$, indicating a zero 
superfluid density.  This means we have a first-layer normal solid as in Ref. 
\onlinecite{tubo1}. The differences between Figs. \ref{enerhe4bajo} and \ref{enerh2bajo}  
can be ascribed basically to the nature of the interparticle interactions: the larger
the potential well, the more clearly  are the limits between the phases.

\begin{figure}
\begin{center}
\includegraphics[width=0.8\linewidth]{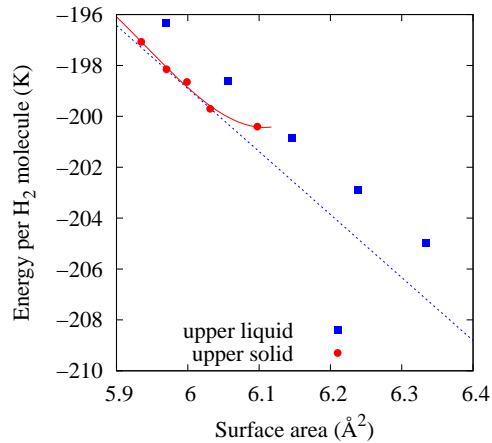}
\caption{Energy per H$_2$ molecule  as a function of the inverse of the total H$_2$ density.  Solid squares,  a two-layer structure with a liquid on top of a first-layer 7-in-a-row solid of density $\rho_s$ = 0.100 \AA$^{-2}$.  Solid circles,  a 15-in-a-row upper solid on top of the same H$_2$ substrate.  The dotted line is the Maxwell construction line between a single layer solid of $\rho$ = 0.0975 \AA$^{-2}$ and a two-layer solid of $\rho$ = 0.166 \AA$^{-2}$. The solid line is a third-order polynomial fit to the data intended exclusively as a guide-to-the eye. }
\label{enerh2alta3}
\end{center}
\end{figure}

Following exactly the same procedure outlined above for the case of $^4$He,  we 
considered 7-in-a-row first-layer H$_2$ solids of different densities and 
deposited additional molecules on top of them.  After that, we kept the 
arrangements with the minimum energy per particle for 
each total (first+second layer) densities.  The second layer densities were 
calculated using as a surface adsorbate a cylinder of radius 9.42 \AA. This is 
the mean distance of the molecules of the second layer to the axis of the tube.  
The minimum energies per molecule corresponded to a first-layer solid of 
$\rho_s$ = 0.100 \AA$^{-2}$ with either a  liquid ($c_2$ = 0 in Eq. \ref{t2}) or 
a (super)solid ($c_2 \neq$ 0) on top.  In Fig. \ref{enerh2alta3}, we show those 
energies as a function of the inverse of the density.  The $x$-range of that 
figure corresponds to the inverse of the experimental densities at which the 
H$_2$ is short of being promoted to a third layer on a flat substrate  
\cite{wiechert}.  The solid data corresponds to a 15-in-a-row phase built in the 
same way that for the first-layer solids.  The dotted line is a double tangent 
Maxwell construction between a single layer 7-in-a-row solid of density $\rho$ = 
0.0975 \AA$^{-2}$ and a two-layer solid of total density $\rho$ = 0.166 
\AA$^{-2}$.  Both densities are within the error bars of the ones for a flat 
second H$_2$ absorbed on graphite \cite{prb2022}. 

In the same way as the second layer of H$_2$ on graphite at low densities,  the 
15-in-a-row H$_2$ on top of a nanotube is also a supersolid in a thin density 
slice. In Fig. \ref{supersolidoh2} we can see (solid squares) the same 
superfluid estimator already described but applied only to the molecules on the 
second layer for a solid of total density $\rho$ = 0.167 \AA$^{-2}$.  For that 
arrangement, the slope
corresponds to a supersolid density of 1.32 $\pm$ 0.05 \%,  larger but of the 
same order of magnitude as the value for the second layer of H$_2$ on graphite 
(0.41 $\pm$ 0.05 \%).  

\section{Discussion}

In this work, we have calculated the phase diagrams of $^4$He and H$_2$ on a 
single carbon nanotube.  We chose the (5,5) one because is one of the thinnest 
experimentally obtained and its narrowness would make it a perfect candidate to 
see the differences between adsorption on a flat substrate and on a quasi-one 
dimensional one.  Here, the dimensionality would be the only factor to take into 
account since the chemical composition of graphite, graphene, and carbon 
nanotubes is identical. 
Our simulation results suggest that the behavior of those species is quite similar on a cylinder and on its flat counterparts. There are, however, some minor differences.  

$^4$He is a supersolid at low densities.  In this, its behavior is similar to 
that of a single sheet of $^4$He on top of graphene and graphite 
\cite{claudiosuper}.  However,  on a tube we have incommensurate structures 
instead  of  the $\sqrt{3} \times \sqrt{3}$ registered supersolid on the flat 
substrates.  In any case, the superfluidity disappears upon $^4$He loading: 
both the incommensurate solid of 0.08 \AA$^{-2}$ on graphene and the 6-in-a-row 
stable solids are normal.  On the other hand, there is a significant difference 
in the promotion to a second layer.  Even though on graphite there is a stable 
liquid in equilibrium with an incommensurate single-sheet solid, the density of 
that liquid is smaller than the one obtained in the present 
work (a total density of 0.163 \AA$^{-2}$ \cite{prl2020} versus the value 0.181 
\AA$^{-2}$ of Table \ref{tab:table1}).   In addition,  we do not see a stable 
second-layer incommensurate solid up to the experimental density corresponding 
to a third layer promotion on graphite \cite{nyeki}.  
%In this, $^4$He on a 
%cylinder is similar bulk $^4$He,  since both are stable liquids even at  T = 
%0.  

On the other hand, the behavior of H$_2$ on the selected nanotube is very 
similar to the one on top of graphene or graphite.  The only difference is 
that, at very low densities, we have an incommensurate normal solid instead of 
the $\sqrt{3} \times \sqrt{3}$ normal structure on the flat adsorbents.  We 
have also a promotion to a second layer supersolid at the same densities as in 
graphite \cite{prb2022}.  The supersolid fraction at the lowest second layer 
densities is also comparable,  albeit slightly larger than the one 
for graphite.   This suggests  that the behavior of H$_2$ depends more strongly 
on the strength of the  H$_2$-H$_2$ interaction than in the dimensionality of the system and that 
it is probably necessary to disrupt the H$_2$ structure to observe new 
H$_2$ phases \cite{prb2023}. Another possible issue could be the quality of 
the empirical potential 
itself, and some alternatives to the Silvera-Goldman expression based on first-principles
have been proposed \cite{h1,h2,h2,h3,h4,h5}, some of them including three-body terms. 
However, the equation of state of solid $hcp$ H$_2$  obtained 
using those interactions is similar to the one calculated using the Silvera and 
Goldman potential \cite{h3} and in reasonably agreement with experiments. 
Moreover, a recent
comparison between neutron scattering data for liquid and solid H$_2$ phases and simulations using the Silvera and Buck potentials
suggests that both interactions are adequate to reproduce the experimental observables \cite{prbsilvera}.   
%This justifies us in the use of that standard interaction to describe the H$_2$ 
%systems. 

The superfluid fraction of the liquid $^4$He layer is one and the one of 
the solid H$_2$ layer is $\sim 0.01$. This large difference is mainly due to 
the different interaction between $^4$He atoms and H$_2$ molecules, the latter 
being much more attractive thus favoring the crystal phases.   
The superfluid fraction of this supersolid phase of H$_2$ may appear to be too 
small to be detected, but similar fractions 
were experimentally measured on the second layer of $^4$He on graphite \cite{kim} 
at temperatures low enough (0.5 K) to make our $T=0$ calculations relevant.    
The experiment~\cite{kim} was  made possible by using a specially designed 
double torsional oscillator able to disentangle the signals coming from the 
superfluid and elastic responses.
%, a technique similar to the one used to detect the presence 
%of absorbents on tubes \cite{tavernarakis,noury}.
On the other hand, a recent calculation \cite{parinello} supports the existence 
of supersolidity in D$_2$ at very high pressures. 

\begin{acknowledgments}
We acknowledge financial support from Ministerio de
Ciencia e Innovación MCIN/AEI/10.13039/501100011033 
(Spain) under Grants No. PID2020-113565GB-C22 and No.
PID2020-113565GB-C21, from Junta de Andaluc\'{\i}a group
PAIDI-205. M.C.G. acknowledges funding from Fondo Europeo de Desarrollo Regional (FEDER) and Consejer\'{\i}a de
Econom\'{\i}a, Conocimiento, Empresas y Universidad de la Junta
de Andaluc\'{\i}a, en marco del programa operativo FEDER  Andaluc\'{\i}a
2014-2020. Objetivo específico 1.2.3. “Fomento y
generaci\'on de conocimiento frontera y de conocimiento orientado a los retos de la sociedad, desarrollo de tecnolog\'{\i}as
emergentes” under Grant No. UPO-1380159, Porcentaje de
cofinanciaci\'on FEDER 80\%, and from  AGAUR-Generalitat de 
Catalunya Grant No. 2021-SGR-01411.
 We also acknowledge
the use of the C3UPO computer facilities at the Universidad
Pablo de Olavide.
\end{acknowledgments}

\bibliography{he4h2tube_boro3}% Produces the bibliography via BibTeX.

\begin{thebibliography}{38}%
\makeatletter
\providecommand \@ifxundefined [1]{%
 \@ifx{#1\undefined}
}%
\providecommand \@ifnum [1]{%
 \ifnum #1\expandafter \@firstoftwo
 \else \expandafter \@secondoftwo
 \fi
}%
\providecommand \@ifx [1]{%
 \ifx #1\expandafter \@firstoftwo
 \else \expandafter \@secondoftwo
 \fi
}%
\providecommand \natexlab [1]{#1}%
\providecommand \enquote  [1]{``#1''}%
\providecommand \bibnamefont  [1]{#1}%
\providecommand \bibfnamefont [1]{#1}%
\providecommand \citenamefont [1]{#1}%
\providecommand \href@noop [0]{\@secondoftwo}%
\providecommand \href [0]{\begingroup \@sanitize@url \@href}%
\providecommand \@href[1]{\@@startlink{#1}\@@href}%
\providecommand \@@href[1]{\endgroup#1\@@endlink}%
\providecommand \@sanitize@url [0]{\catcode `\\12\catcode `\$12\catcode
  `\&12\catcode `\#12\catcode `\^12\catcode `\_12\catcode `\%12\relax}%
\providecommand \@@startlink[1]{}%
\providecommand \@@endlink[0]{}%
\providecommand \url  [0]{\begingroup\@sanitize@url \@url }%
\providecommand \@url [1]{\endgroup\@href {#1}{\urlprefix }}%
\providecommand \urlprefix  [0]{URL }%
\providecommand \Eprint [0]{\href }%
\providecommand \doibase [0]{https://doi.org/}%
\providecommand \selectlanguage [0]{\@gobble}%
\providecommand \bibinfo  [0]{\@secondoftwo}%
\providecommand \bibfield  [0]{\@secondoftwo}%
\providecommand \translation [1]{[#1]}%
\providecommand \BibitemOpen [0]{}%
\providecommand \bibitemStop [0]{}%
\providecommand \bibitemNoStop [0]{.\EOS\space}%
\providecommand \EOS [0]{\spacefactor3000\relax}%
\providecommand \BibitemShut  [1]{\csname bibitem#1\endcsname}%
\let\auto@bib@innerbib\@empty
%</preamble>
\bibitem [{\citenamefont {Iijima}(1991)}]{iijima}%
  \BibitemOpen
  \bibfield  {author} {\bibinfo {author} {\bibfnamefont {S.}~\bibnamefont
  {Iijima}},\ }\bibfield  {title} {\bibinfo {title} {Helical microtubules of
  graphitic carbon},\ }\href@noop {} {\bibfield  {journal} {\bibinfo  {journal}
  {Nature}\ }\textbf {\bibinfo {volume} {354}},\ \bibinfo {pages} {56}
  (\bibinfo {year} {1991})}\BibitemShut {NoStop}%
\bibitem [{\citenamefont {Wang}\ \emph {et~al.}(2010)\citenamefont {Wang},
  \citenamefont {{\relax J. Wei, P. Morse, J. G. Dash and O. E. Vilches}},\
  and\ \citenamefont {Cobden}}]{cobden1}%
  \BibitemOpen
  \bibfield  {author} {\bibinfo {author} {\bibfnamefont {Z.}~\bibnamefont
  {Wang}}, \bibinfo {author} {\bibnamefont {{\relax J. Wei, P. Morse, J. G.
  Dash and O. E. Vilches}}},\ and\ \bibinfo {author} {\bibfnamefont {D.~H.}\
  \bibnamefont {Cobden}},\ }\bibfield  {title} {\bibinfo {title} {Phase
  transitions of adsorbed atoms on the surface of a carbon nanotube},\ }\href
  {https://doi.org/10.1126/science.1182507} {\bibfield  {journal} {\bibinfo
  {journal} {Science}\ }\textbf {\bibinfo {volume} {327}},\ \bibinfo {pages}
  {552} (\bibinfo {year} {2010})}\BibitemShut {NoStop}%
\bibitem [{\citenamefont {Lee}\ \emph {et~al.}(2012)\citenamefont {Lee},
  \citenamefont {{\relax O. E. Vilches, Z. Wang, E. Fredrickson, P. Morse, R.
  Roy and D. Dzyubenko}},\ and\ \citenamefont {Cobden}}]{cobden2}%
  \BibitemOpen
  \bibfield  {author} {\bibinfo {author} {\bibfnamefont {H.}~\bibnamefont
  {Lee}}, \bibinfo {author} {\bibnamefont {{\relax O. E. Vilches, Z. Wang, E.
  Fredrickson, P. Morse, R. Roy and D. Dzyubenko}}},\ and\ \bibinfo {author}
  {\bibfnamefont {D.~H.}\ \bibnamefont {Cobden}},\ }\bibfield  {title}
  {\bibinfo {title} {{Kr and $^4$He Adsorption on Individual Suspended
  Single-Walled Carbon Nanotubes }},\ }\href
  {https://doi.org/10.1007/s10909-012-0642-3} {\bibfield  {journal} {\bibinfo
  {journal} {J. Low Temp. Phys.}\ }\textbf {\bibinfo {volume} {169}},\ \bibinfo
  {pages} {338} (\bibinfo {year} {2012})}\BibitemShut {NoStop}%
\bibitem [{\citenamefont {Chiu}\ \emph {et~al.}(2008)\citenamefont {Chiu},
  \citenamefont {{\relax P. Hung and H.W. Ch. Postma}},\ and\ \citenamefont
  {Bockrath}}]{chiu}%
  \BibitemOpen
  \bibfield  {author} {\bibinfo {author} {\bibfnamefont {H.}~\bibnamefont
  {Chiu}}, \bibinfo {author} {\bibnamefont {{\relax P. Hung and H.W. Ch.
  Postma}}},\ and\ \bibinfo {author} {\bibfnamefont {M.}~\bibnamefont
  {Bockrath}},\ }\bibfield  {title} {\bibinfo {title} {Atomic-scale mass
  sensing using carbon nanotube resonators},\ }\href
  {https://doi.org/10.1021/nl802181c} {\bibfield  {journal} {\bibinfo
  {journal} {Nano Lett.}\ }\textbf {\bibinfo {volume} {8}},\ \bibinfo {pages}
  {4342} (\bibinfo {year} {2008})}\BibitemShut {NoStop}%
\bibitem [{\citenamefont {Chaste}\ \emph {et~al.}(2012)\citenamefont {Chaste},
  \citenamefont {{\relax A. Eichler, J. Moser, G. Ceballos and R. Rurali}},\
  and\ \citenamefont {Bachtold}}]{chaste1}%
  \BibitemOpen
  \bibfield  {author} {\bibinfo {author} {\bibfnamefont {J.}~\bibnamefont
  {Chaste}}, \bibinfo {author} {\bibnamefont {{\relax A. Eichler, J. Moser, G.
  Ceballos and R. Rurali}}},\ and\ \bibinfo {author} {\bibfnamefont
  {A.}~\bibnamefont {Bachtold}},\ }\bibfield  {title} {\bibinfo {title} {A
  nanomechanical mass sensor with yoctogram resolution},\ }\href
  {https://doi.org/10.1038/nnano.2012.42} {\bibfield  {journal} {\bibinfo
  {journal} {Nature Nanotech.}\ }\textbf {\bibinfo {volume} {7}},\ \bibinfo
  {pages} {301} (\bibinfo {year} {2012})}\BibitemShut {NoStop}%
\bibitem [{\citenamefont {Tavernarakis}\ \emph {et~al.}(2014)\citenamefont
  {Tavernarakis}, \citenamefont {Chaste}, \citenamefont {Eichler},
  \citenamefont {Ceballos}, \citenamefont {Gordillo}, \citenamefont {Boronat},\
  and\ \citenamefont {Bachtold}}]{tavernarakis}%
  \BibitemOpen
  \bibfield  {author} {\bibinfo {author} {\bibfnamefont {A.}~\bibnamefont
  {Tavernarakis}}, \bibinfo {author} {\bibfnamefont {J.}~\bibnamefont
  {Chaste}}, \bibinfo {author} {\bibfnamefont {A.}~\bibnamefont {Eichler}},
  \bibinfo {author} {\bibfnamefont {G.}~\bibnamefont {Ceballos}}, \bibinfo
  {author} {\bibfnamefont {M.~C.}\ \bibnamefont {Gordillo}}, \bibinfo {author}
  {\bibfnamefont {J.}~\bibnamefont {Boronat}},\ and\ \bibinfo {author}
  {\bibfnamefont {A.}~\bibnamefont {Bachtold}},\ }\bibfield  {title} {\bibinfo
  {title} {Atomic monolayer deposition on the surface of nanotube mechanical
  resonators},\ }\href {https://doi.org/10.1103/PhysRevLett.112.196103}
  {\bibfield  {journal} {\bibinfo  {journal} {Phys. Rev. Lett.}\ }\textbf
  {\bibinfo {volume} {112}},\ \bibinfo {pages} {196103} (\bibinfo {year}
  {2014})}\BibitemShut {NoStop}%
\bibitem [{\citenamefont {Noury}\ \emph {et~al.}(2019)\citenamefont {Noury},
  \citenamefont {Vergara-Cruz}, \citenamefont {Morfin}, \citenamefont
  {Pla\ifmmode~\mbox{\c{c}}\else \c{c}\fi{}ais}, \citenamefont {Gordillo},
  \citenamefont {Boronat}, \citenamefont {Balibar},\ and\ \citenamefont
  {Bachtold}}]{noury}%
  \BibitemOpen
  \bibfield  {author} {\bibinfo {author} {\bibfnamefont {A.}~\bibnamefont
  {Noury}}, \bibinfo {author} {\bibfnamefont {J.}~\bibnamefont {Vergara-Cruz}},
  \bibinfo {author} {\bibfnamefont {P.}~\bibnamefont {Morfin}}, \bibinfo
  {author} {\bibfnamefont {B.}~\bibnamefont {Pla\ifmmode~\mbox{\c{c}}\else
  \c{c}\fi{}ais}}, \bibinfo {author} {\bibfnamefont {M.~C.}\ \bibnamefont
  {Gordillo}}, \bibinfo {author} {\bibfnamefont {J.}~\bibnamefont {Boronat}},
  \bibinfo {author} {\bibfnamefont {S.}~\bibnamefont {Balibar}},\ and\ \bibinfo
  {author} {\bibfnamefont {A.}~\bibnamefont {Bachtold}},\ }\bibfield  {title}
  {\bibinfo {title} {Layering transition in superfluid helium adsorbed on a
  carbon nanotube mechanical resonator},\ }\href
  {https://doi.org/10.1103/PhysRevLett.122.165301} {\bibfield  {journal}
  {\bibinfo  {journal} {Phys. Rev. Lett.}\ }\textbf {\bibinfo {volume} {122}},\
  \bibinfo {pages} {165301} (\bibinfo {year} {2019})}\BibitemShut {NoStop}%
\bibitem [{\citenamefont {Todoshchenko}\ \emph {et~al.}(2022)\citenamefont
  {Todoshchenko}, \citenamefont {{\relax M. Kamada, J.P. Kaikkonen, Y. Liao, A.
  Savin, M. Will, E. Sergeicheva, T. S. Abhilash and E. Kauppinen}},\ and\
  \citenamefont {Hakonen}}]{todoshchenko}%
  \BibitemOpen
  \bibfield  {author} {\bibinfo {author} {\bibfnamefont {I.}~\bibnamefont
  {Todoshchenko}}, \bibinfo {author} {\bibnamefont {{\relax M. Kamada, J.P.
  Kaikkonen, Y. Liao, A. Savin, M. Will, E. Sergeicheva, T. S. Abhilash and E.
  Kauppinen}}},\ and\ \bibinfo {author} {\bibfnamefont {P.}~\bibnamefont
  {Hakonen}},\ }\bibfield  {title} {\bibinfo {title} {{Topologically-imposed
  vacancies and mobile solid $^3$He on carbon nanotube}},\ }\href
  {https://doi.org/10.1038/s41467-022-33539-8} {\bibfield  {journal} {\bibinfo
  {journal} {Nat. Commun.}\ }\textbf {\bibinfo {volume} {13}},\ \bibinfo
  {pages} {5873} (\bibinfo {year} {2022})}\BibitemShut {NoStop}%
\bibitem [{\citenamefont {Nyeki}\ \emph {et~al.}(2017)\citenamefont {Nyeki},
  \citenamefont {{\relax A. Phillis, A. Ho, D. Lee, P. Coleman, J. Parpia and
  B. Cowan}},\ and\ \citenamefont {Saunders}}]{nyeki}%
  \BibitemOpen
  \bibfield  {author} {\bibinfo {author} {\bibfnamefont {J.}~\bibnamefont
  {Nyeki}}, \bibinfo {author} {\bibnamefont {{\relax A. Phillis, A. Ho, D. Lee,
  P. Coleman, J. Parpia and B. Cowan}}},\ and\ \bibinfo {author} {\bibfnamefont
  {J.}~\bibnamefont {Saunders}},\ }\bibfield  {title} {\bibinfo {title}
  {{Intertwined superfluid and density wave order in two-dimensional$^4$He}},\
  }\href {https://doi.org/10.1038/nphys4023} {\bibfield  {journal} {\bibinfo
  {journal} {Nat. Phys.}\ }\textbf {\bibinfo {volume} {13}},\ \bibinfo {pages}
  {455} (\bibinfo {year} {2017})}\BibitemShut {NoStop}%
\bibitem [{\citenamefont {Choi}\ \emph {et~al.}(2021)\citenamefont {Choi},
  \citenamefont {{\relax A.A. Zadorozhko, Jeakyung. Choi}},\ and\ \citenamefont
  {Kim}}]{kim}%
  \BibitemOpen
  \bibfield  {author} {\bibinfo {author} {\bibfnamefont {J.}~\bibnamefont
  {Choi}}, \bibinfo {author} {\bibnamefont {{\relax A.A. Zadorozhko, Jeakyung.
  Choi}}},\ and\ \bibinfo {author} {\bibfnamefont {E.}~\bibnamefont {Kim}},\
  }\bibfield  {title} {\bibinfo {title} {{Spatially modulated Superfluid State
  in two-dimensional $^4$He films}},\ }\href@noop {} {\bibfield  {journal}
  {\bibinfo  {journal} {Phys.\ Rev.\ Lett.}\ }\textbf {\bibinfo {volume}
  {127}},\ \bibinfo {pages} {135301} (\bibinfo {year} {2021})}\BibitemShut
  {NoStop}%
\bibitem [{\citenamefont {Nakamura}\ \emph {et~al.}(2016)\citenamefont
  {Nakamura}, \citenamefont {Matsui}, \citenamefont {Matsui},\ and\
  \citenamefont {Fukuyama}}]{fukuyama}%
  \BibitemOpen
  \bibfield  {author} {\bibinfo {author} {\bibfnamefont {S.}~\bibnamefont
  {Nakamura}}, \bibinfo {author} {\bibfnamefont {K.}~\bibnamefont {Matsui}},
  \bibinfo {author} {\bibfnamefont {T.}~\bibnamefont {Matsui}},\ and\ \bibinfo
  {author} {\bibfnamefont {H.}~\bibnamefont {Fukuyama}},\ }\bibfield  {title}
  {\bibinfo {title} {Possible quantum liquid crystal phases of helium
  monolayers},\ }\href {https://doi.org/10.1103/PhysRevB.94.180501} {\bibfield
  {journal} {\bibinfo  {journal} {Phys. Rev. B}\ }\textbf {\bibinfo {volume}
  {94}},\ \bibinfo {pages} {180501(R)} (\bibinfo {year} {2016})}\BibitemShut
  {NoStop}%
\bibitem [{\citenamefont {Gordillo}\ \emph {et~al.}(2011)\citenamefont
  {Gordillo}, \citenamefont {Cazorla},\ and\ \citenamefont
  {Boronat}}]{claudiosuper}%
  \BibitemOpen
  \bibfield  {author} {\bibinfo {author} {\bibfnamefont {M.~C.}\ \bibnamefont
  {Gordillo}}, \bibinfo {author} {\bibfnamefont {C.}~\bibnamefont {Cazorla}},\
  and\ \bibinfo {author} {\bibfnamefont {J.}~\bibnamefont {Boronat}},\
  }\bibfield  {title} {\bibinfo {title} {Supersolidity in quantum films
  adsorbed on graphene and graphite},\ }\href
  {https://doi.org/10.1103/PhysRevB.83.121406} {\bibfield  {journal} {\bibinfo
  {journal} {Phys. Rev. B}\ }\textbf {\bibinfo {volume} {83}},\ \bibinfo
  {pages} {121406(R)} (\bibinfo {year} {2011})}\BibitemShut {NoStop}%
\bibitem [{\citenamefont {Gordillo}\ and\ \citenamefont
  {Boronat}(2020)}]{prl2020}%
  \BibitemOpen
  \bibfield  {author} {\bibinfo {author} {\bibfnamefont {M.~C.}\ \bibnamefont
  {Gordillo}}\ and\ \bibinfo {author} {\bibfnamefont {J.}~\bibnamefont
  {Boronat}},\ }\bibfield  {title} {\bibinfo {title} {Superfluid and supersolid
  phases of $^{4}\mathrm{He}$ on the second layer of graphite},\ }\href
  {https://doi.org/10.1103/PhysRevLett.124.205301} {\bibfield  {journal}
  {\bibinfo  {journal} {Phys. Rev. Lett.}\ }\textbf {\bibinfo {volume} {124}},\
  \bibinfo {pages} {205301} (\bibinfo {year} {2020})}\BibitemShut {NoStop}%
\bibitem [{\citenamefont {Gordillo}\ and\ \citenamefont
  {Boronat}(2022)}]{prb2022}%
  \BibitemOpen
  \bibfield  {author} {\bibinfo {author} {\bibfnamefont {M.~C.}\ \bibnamefont
  {Gordillo}}\ and\ \bibinfo {author} {\bibfnamefont {J.}~\bibnamefont
  {Boronat}},\ }\bibfield  {title} {\bibinfo {title} {Supersolidity in the
  second layer of para-${\mathrm{h}}_{2}$ adsorbed on graphite},\ }\href
  {https://doi.org/10.1103/PhysRevB.105.094501} {\bibfield  {journal} {\bibinfo
   {journal} {Phys. Rev. B}\ }\textbf {\bibinfo {volume} {105}},\ \bibinfo
  {pages} {094501} (\bibinfo {year} {2022})}\BibitemShut {NoStop}%
\bibitem [{\citenamefont {Gordillo}\ and\ \citenamefont
  {Boronat}(2012{\natexlab{a}})}]{tubo2}%
  \BibitemOpen
  \bibfield  {author} {\bibinfo {author} {\bibfnamefont {M.~C.}\ \bibnamefont
  {Gordillo}}\ and\ \bibinfo {author} {\bibfnamefont {J.}~\bibnamefont
  {Boronat}},\ }\bibfield  {title} {\bibinfo {title} {{${}^{\mathbf{4}}$He
  adsorbed outside a single carbon nanotube}},\ }\href
  {https://doi.org/10.1103/PhysRevB.86.165409} {\bibfield  {journal} {\bibinfo
  {journal} {Phys. Rev. B}\ }\textbf {\bibinfo {volume} {86}},\ \bibinfo
  {pages} {165409} (\bibinfo {year} {2012}{\natexlab{a}})}\BibitemShut
  {NoStop}%
\bibitem [{\citenamefont {Gordillo}\ and\ \citenamefont
  {Boronat}(2011)}]{tubo1}%
  \BibitemOpen
  \bibfield  {author} {\bibinfo {author} {\bibfnamefont {M.~C.}\ \bibnamefont
  {Gordillo}}\ and\ \bibinfo {author} {\bibfnamefont {J.}~\bibnamefont
  {Boronat}},\ }\bibfield  {title} {\bibinfo {title} {{Phase transitions of
  ${\mathrm{H}}_{2}$ adsorbed on the surface of single carbon nanotubes}},\
  }\href {https://doi.org/10.1103/PhysRevB.84.033406} {\bibfield  {journal}
  {\bibinfo  {journal} {Phys. Rev. B}\ }\textbf {\bibinfo {volume} {84}},\
  \bibinfo {pages} {033406} (\bibinfo {year} {2011})}\BibitemShut {NoStop}%
\bibitem [{\citenamefont {Carlos}\ and\ \citenamefont
  {Cole}(1980)}]{carlosandcole}%
  \BibitemOpen
  \bibfield  {author} {\bibinfo {author} {\bibfnamefont {W.}~\bibnamefont
  {Carlos}}\ and\ \bibinfo {author} {\bibfnamefont {M.~W.}\ \bibnamefont
  {Cole}},\ }\bibfield  {title} {\bibinfo {title} {Interaction between a he
  atom and a graphite surface},\ }\href
  {https://doi.org/10.1016/0039-6028(80)90090-4} {\bibfield  {journal}
  {\bibinfo  {journal} {Surf. Sci.}\ }\textbf {\bibinfo {volume} {91}},\
  \bibinfo {pages} {339} (\bibinfo {year} {1980})}\BibitemShut {NoStop}%
\bibitem [{\citenamefont {Stan}\ and\ \citenamefont {Cole}(1998)}]{coleh2}%
  \BibitemOpen
  \bibfield  {author} {\bibinfo {author} {\bibfnamefont {G.}~\bibnamefont
  {Stan}}\ and\ \bibinfo {author} {\bibfnamefont {M.~W.}\ \bibnamefont
  {Cole}},\ }\bibfield  {title} {\bibinfo {title} {Hydrogen adsorption in
  nanotubes},\ }\href {https://doi.org/10.1023/A:1022558800315} {\bibfield
  {journal} {\bibinfo  {journal} {J. Low Temp. Phys.}\ }\textbf {\bibinfo
  {volume} {110}},\ \bibinfo {pages} {539} (\bibinfo {year}
  {1998})}\BibitemShut {NoStop}%
\bibitem [{\citenamefont {R.A.~Aziz}\ and\ \citenamefont {Wong}(1987)}]{aziz}%
  \BibitemOpen
  \bibfield  {author} {\bibinfo {author} {\bibfnamefont {F.~M.}\ \bibnamefont
  {R.A.~Aziz}}\ and\ \bibinfo {author} {\bibfnamefont {C.~K.}\ \bibnamefont
  {Wong}},\ }\bibfield  {title} {\bibinfo {title} {{A new determination of the
  ground state interatomic potential for He$_2$}},\ }\href@noop {} {\bibfield
  {journal} {\bibinfo  {journal} {Mol. Phys.}\ }\textbf {\bibinfo {volume}
  {61}},\ \bibinfo {pages} {1487} (\bibinfo {year} {1987})}\BibitemShut
  {NoStop}%
\bibitem [{\citenamefont {Silvera}\ and\ \citenamefont
  {Goldman}(1978)}]{silvera}%
  \BibitemOpen
  \bibfield  {author} {\bibinfo {author} {\bibfnamefont {I.~F.}\ \bibnamefont
  {Silvera}}\ and\ \bibinfo {author} {\bibfnamefont {V.~V.}\ \bibnamefont
  {Goldman}},\ }\bibfield  {title} {\bibinfo {title} {{The isotropic
  intermolecular potential for H$_2$ and D$_2$ in the solid and gas phases}},\
  }\href@noop {} {\bibfield  {journal} {\bibinfo  {journal} {J. Chem. Phys.}\
  }\textbf {\bibinfo {volume} {69}},\ \bibinfo {pages} {4209} (\bibinfo {year}
  {1978})}\BibitemShut {NoStop}%
\bibitem [{\citenamefont {Gordillo}\ and\ \citenamefont
  {Boronat}(2009)}]{prl2009}%
  \BibitemOpen
  \bibfield  {author} {\bibinfo {author} {\bibfnamefont {M.~C.}\ \bibnamefont
  {Gordillo}}\ and\ \bibinfo {author} {\bibfnamefont {J.}~\bibnamefont
  {Boronat}},\ }\bibfield  {title} {\bibinfo {title} {{$^4$He on a single
  graphene sheet}},\ }\href@noop {} {\bibfield  {journal} {\bibinfo  {journal}
  {Phys.\ Rev.\ Lett.}\ }\textbf {\bibinfo {volume} {102}},\ \bibinfo {pages}
  {085303} (\bibinfo {year} {2009})}\BibitemShut {NoStop}%
\bibitem [{\citenamefont {Gordillo}\ and\ \citenamefont
  {Boronat}(2010)}]{prb2010}%
  \BibitemOpen
  \bibfield  {author} {\bibinfo {author} {\bibfnamefont {M.~C.}\ \bibnamefont
  {Gordillo}}\ and\ \bibinfo {author} {\bibfnamefont {J.}~\bibnamefont
  {Boronat}},\ }\bibfield  {title} {\bibinfo {title} {{Phase diagram of H$_2$
  adsorbed on graphene}},\ }\href@noop {} {\bibfield  {journal} {\bibinfo
  {journal} {Phys.\ Rev.\ B}\ }\textbf {\bibinfo {volume} {81}},\ \bibinfo
  {pages} {155435} (\bibinfo {year} {2010})}\BibitemShut {NoStop}%
\bibitem [{\citenamefont {Crowell}\ and\ \citenamefont
  {Reppy}(1993)}]{crowell1}%
  \BibitemOpen
  \bibfield  {author} {\bibinfo {author} {\bibfnamefont {P.~A.}\ \bibnamefont
  {Crowell}}\ and\ \bibinfo {author} {\bibfnamefont {J.~D.}\ \bibnamefont
  {Reppy}},\ }\bibfield  {title} {\bibinfo {title} {Reentrant superfluidity in
  $^{4}\mathrm{He}$ films adsorbed on graphite},\ }\href
  {https://doi.org/10.1103/PhysRevLett.70.3291} {\bibfield  {journal} {\bibinfo
   {journal} {Phys. Rev. Lett.}\ }\textbf {\bibinfo {volume} {70}},\ \bibinfo
  {pages} {3291} (\bibinfo {year} {1993})}\BibitemShut {NoStop}%
\bibitem [{\citenamefont {Crowell}\ and\ \citenamefont
  {Reppy}(1996)}]{crowell2}%
  \BibitemOpen
  \bibfield  {author} {\bibinfo {author} {\bibfnamefont {P.~A.}\ \bibnamefont
  {Crowell}}\ and\ \bibinfo {author} {\bibfnamefont {J.~D.}\ \bibnamefont
  {Reppy}},\ }\bibfield  {title} {\bibinfo {title} {Superfluidity and film
  structure in $^{4}\mathrm{He}$ adsorbed on graphite},\ }\href
  {https://doi.org/10.1103/PhysRevB.53.2701} {\bibfield  {journal} {\bibinfo
  {journal} {Phys. Rev. B}\ }\textbf {\bibinfo {volume} {53}},\ \bibinfo
  {pages} {2701} (\bibinfo {year} {1996})}\BibitemShut {NoStop}%
\bibitem [{\citenamefont {Greywall}\ and\ \citenamefont
  {Busch}(1991)}]{greywall1}%
  \BibitemOpen
  \bibfield  {author} {\bibinfo {author} {\bibfnamefont {D.~S.}\ \bibnamefont
  {Greywall}}\ and\ \bibinfo {author} {\bibfnamefont {P.~A.}\ \bibnamefont
  {Busch}},\ }\bibfield  {title} {\bibinfo {title} {Heat capacity of fluid
  monolayers of $^{4}\mathrm{He}$},\ }\href
  {https://doi.org/10.1103/PhysRevLett.67.3535} {\bibfield  {journal} {\bibinfo
   {journal} {Phys. Rev. Lett.}\ }\textbf {\bibinfo {volume} {67}},\ \bibinfo
  {pages} {3535} (\bibinfo {year} {1991})}\BibitemShut {NoStop}%
\bibitem [{\citenamefont {Greywall}(1993)}]{greywall2}%
  \BibitemOpen
  \bibfield  {author} {\bibinfo {author} {\bibfnamefont {D.~S.}\ \bibnamefont
  {Greywall}},\ }\bibfield  {title} {\bibinfo {title} {Heat capacity and the
  commensurate-incommensurate transition of $^{4}\mathrm{He}$ adsorbed on
  graphite},\ }\href {https://doi.org/10.1103/PhysRevB.47.309} {\bibfield
  {journal} {\bibinfo  {journal} {Phys. Rev. B}\ }\textbf {\bibinfo {volume}
  {47}},\ \bibinfo {pages} {309} (\bibinfo {year} {1993})}\BibitemShut
  {NoStop}%
\bibitem [{\citenamefont {Wiechert}(1991)}]{wiechert}%
  \BibitemOpen
  \bibfield  {author} {\bibinfo {author} {\bibfnamefont {H.}~\bibnamefont
  {Wiechert}},\ }in\ \href@noop {} {\emph {\bibinfo {booktitle} {Excitations in
  Two-Dimensional and Three-Dimensional Quantum Fluid}}},\ \bibinfo {editor}
  {edited by\ \bibinfo {editor} {\bibfnamefont {A.}~\bibnamefont {Wyatt}}\ and\
  \bibinfo {editor} {\bibfnamefont {H.}~\bibnamefont {Lauter}}}\ (\bibinfo
  {publisher} {Plenum},\ \bibinfo {address} {New York},\ \bibinfo {year}
  {1991})\BibitemShut {NoStop}%
\bibitem [{\citenamefont {Gordillo}\ and\ \citenamefont
  {Boronat}(2013)}]{prbsecondh2}%
  \BibitemOpen
  \bibfield  {author} {\bibinfo {author} {\bibfnamefont {M.~C.}\ \bibnamefont
  {Gordillo}}\ and\ \bibinfo {author} {\bibfnamefont {J.}~\bibnamefont
  {Boronat}},\ }\bibfield  {title} {\bibinfo {title} {Second layer of h${}_{2}$
  and d${}_{2}$ adsorbed on graphene},\ }\href
  {https://doi.org/10.1103/PhysRevB.87.165403} {\bibfield  {journal} {\bibinfo
  {journal} {Phys. Rev. B}\ }\textbf {\bibinfo {volume} {87}},\ \bibinfo
  {pages} {165403} (\bibinfo {year} {2013})}\BibitemShut {NoStop}%
\bibitem [{\citenamefont {Zhang}\ \emph {et~al.}(1995)\citenamefont {Zhang},
  \citenamefont {{\relax and N. Kawashima, J. Carlson}},\ and\ \citenamefont
  {Gubernatis}}]{gubernatis}%
  \BibitemOpen
  \bibfield  {author} {\bibinfo {author} {\bibfnamefont {S.}~\bibnamefont
  {Zhang}}, \bibinfo {author} {\bibnamefont {{\relax and N. Kawashima, J.
  Carlson}}},\ and\ \bibinfo {author} {\bibfnamefont {J.~E.}\ \bibnamefont
  {Gubernatis}},\ }\bibfield  {title} {\bibinfo {title} {Quantum simulations of
  the superfluid-insulator transition for two-dimensional, disordered,
  hard-core bosons},\ }\href {https://doi.org/10.1103/PhysRevLett.74.1500}
  {\bibfield  {journal} {\bibinfo  {journal} {Phys. Rev. Lett.}\ }\textbf
  {\bibinfo {volume} {74}},\ \bibinfo {pages} {1500} (\bibinfo {year}
  {1995})}\BibitemShut {NoStop}%
\bibitem [{\citenamefont {Gordillo}\ and\ \citenamefont
  {Boronat}(2012{\natexlab{b}})}]{prbsecondhe4}%
  \BibitemOpen
  \bibfield  {author} {\bibinfo {author} {\bibfnamefont {M.~C.}\ \bibnamefont
  {Gordillo}}\ and\ \bibinfo {author} {\bibfnamefont {J.}~\bibnamefont
  {Boronat}},\ }\bibfield  {title} {\bibinfo {title} {Zero-temperature phase
  diagram of the second layer of ${}^{4}$he adsorbed on graphene},\ }\href
  {https://doi.org/10.1103/PhysRevB.85.195457} {\bibfield  {journal} {\bibinfo
  {journal} {Phys. Rev. B}\ }\textbf {\bibinfo {volume} {85}},\ \bibinfo
  {pages} {195457} (\bibinfo {year} {2012}{\natexlab{b}})}\BibitemShut
  {NoStop}%
\bibitem [{\citenamefont {Gordillo}\ and\ \citenamefont
  {Boronat}(2023)}]{prb2023}%
  \BibitemOpen
  \bibfield  {author} {\bibinfo {author} {\bibfnamefont {M.~C.}\ \bibnamefont
  {Gordillo}}\ and\ \bibinfo {author} {\bibfnamefont {J.}~\bibnamefont
  {Boronat}},\ }\bibfield  {title} {\bibinfo {title} {{${\mathrm{H}}_{2}$
  superglass on an amorphous carbon substrate}},\ }\href
  {https://doi.org/10.1103/PhysRevB.107.L060505} {\bibfield  {journal}
  {\bibinfo  {journal} {Phys. Rev. B}\ }\textbf {\bibinfo {volume} {107}},\
  \bibinfo {pages} {L060505} (\bibinfo {year} {2023})}\BibitemShut {NoStop}%
\bibitem [{\citenamefont {Faruk}\ \emph {et~al.}(2014)\citenamefont {Faruk},
  \citenamefont {{\relax M. Schmidt, H. Li and R. J. Le Roy}},\ and\
  \citenamefont {Roy}}]{h1}%
  \BibitemOpen
  \bibfield  {author} {\bibinfo {author} {\bibfnamefont {N.}~\bibnamefont
  {Faruk}}, \bibinfo {author} {\bibnamefont {{\relax M. Schmidt, H. Li and R.
  J. Le Roy}}},\ and\ \bibinfo {author} {\bibfnamefont {P.}~\bibnamefont
  {Roy}},\ }\bibfield  {title} {\bibinfo {title} {First-principles prediction
  of the raman shifts in parahydrogen clusters},\ }\href
  {https://doi.org/10.1063/1.48852755} {\bibfield  {journal} {\bibinfo
  {journal} {J. Chem. Phys.}\ }\textbf {\bibinfo {volume} {141}},\ \bibinfo
  {pages} {014310} (\bibinfo {year} {2014})}\BibitemShut {NoStop}%
\bibitem [{\citenamefont {Schmidt}\ \emph {et~al.}(2015)\citenamefont
  {Schmidt}, \citenamefont {{\relax J. M. Fern\'ndez, N. Faruk, M. Nooijen, R.
  J. Le Roy, J.H. Morilla, G. Tejeda and S. Montero}},\ and\ \citenamefont
  {Roy}}]{h2}%
  \BibitemOpen
  \bibfield  {author} {\bibinfo {author} {\bibfnamefont {M.}~\bibnamefont
  {Schmidt}}, \bibinfo {author} {\bibnamefont {{\relax J. M. Fern\'ndez, N.
  Faruk, M. Nooijen, R. J. Le Roy, J.H. Morilla, G. Tejeda and S. Montero}}},\
  and\ \bibinfo {author} {\bibfnamefont {P.}~\bibnamefont {Roy}},\ }\bibfield
  {title} {\bibinfo {title} {Raman vibrational shifts of small clusters of
  hydrogen isotopologues},\ }\href {https://doi.org/10.1021/acs.jpca.5b08852}
  {\bibfield  {journal} {\bibinfo  {journal} {J. Phys. Chem. A}\ }\textbf
  {\bibinfo {volume} {119}},\ \bibinfo {pages} {12551} (\bibinfo {year}
  {2015})}\BibitemShut {NoStop}%
\bibitem [{\citenamefont {Ibrahim}\ \emph {et~al.}(2019)\citenamefont
  {Ibrahim}, \citenamefont {{\relax L. Wang, T. Halverson and R. J. Le Roy}},\
  and\ \citenamefont {Roy}}]{h3}%
  \BibitemOpen
  \bibfield  {author} {\bibinfo {author} {\bibfnamefont {A.}~\bibnamefont
  {Ibrahim}}, \bibinfo {author} {\bibnamefont {{\relax L. Wang, T. Halverson
  and R. J. Le Roy}}},\ and\ \bibinfo {author} {\bibfnamefont {P.}~\bibnamefont
  {Roy}},\ }\bibfield  {title} {\bibinfo {title} {Equation of state and first
  principles prediction of the vibrational matrix shift of solid
  parahydrogen},\ }\href {https://doi.org/10.1063/1.5131329} {\bibfield
  {journal} {\bibinfo  {journal} {J. Chem. Phys.}\ }\textbf {\bibinfo {volume}
  {151}},\ \bibinfo {pages} {244501} (\bibinfo {year} {2019})}\BibitemShut
  {NoStop}%
\bibitem [{\citenamefont {Ibrahim}\ and\ \citenamefont
  {Roy}(2022{\natexlab{a}})}]{h4}%
  \BibitemOpen
  \bibfield  {author} {\bibinfo {author} {\bibfnamefont {A.}~\bibnamefont
  {Ibrahim}}\ and\ \bibinfo {author} {\bibfnamefont {P.}~\bibnamefont {Roy}},\
  }\bibfield  {title} {\bibinfo {title} {Three-body potential energy surface
  for para-hydrogen},\ }\href {https://doi.org/10.1063/5.0076494} {\bibfield
  {journal} {\bibinfo  {journal} {J. Chem. Phys.}\ }\textbf {\bibinfo {volume}
  {153}},\ \bibinfo {pages} {044301} (\bibinfo {year}
  {2022}{\natexlab{a}})}\BibitemShut {NoStop}%
\bibitem [{\citenamefont {Ibrahim}\ and\ \citenamefont
  {Roy}(2022{\natexlab{b}})}]{h5}%
  \BibitemOpen
  \bibfield  {author} {\bibinfo {author} {\bibfnamefont {A.}~\bibnamefont
  {Ibrahim}}\ and\ \bibinfo {author} {\bibfnamefont {P.}~\bibnamefont {Roy}},\
  }\bibfield  {title} {\bibinfo {title} {Equation of state of solid
  parahydrogen using ab initio two-body and three-body interaction
  potentials},\ }\href {https://doi.org/10.1063/5.0120169} {\bibfield
  {journal} {\bibinfo  {journal} {J. Chem. Phys.}\ }\textbf {\bibinfo {volume}
  {157}},\ \bibinfo {pages} {174503} (\bibinfo {year}
  {2022}{\natexlab{b}})}\BibitemShut {NoStop}%
\bibitem [{\citenamefont {Prisk}\ \emph {et~al.}(2023)\citenamefont {Prisk},
  \citenamefont {Azuah}, \citenamefont {Abernathy}, \citenamefont {Granroth},
  \citenamefont {Sherline}, \citenamefont {Sokol}, \citenamefont {Hu},\ and\
  \citenamefont {Boninsegni}}]{prbsilvera}%
  \BibitemOpen
  \bibfield  {author} {\bibinfo {author} {\bibfnamefont {T.~R.}\ \bibnamefont
  {Prisk}}, \bibinfo {author} {\bibfnamefont {R.~T.}\ \bibnamefont {Azuah}},
  \bibinfo {author} {\bibfnamefont {D.~L.}\ \bibnamefont {Abernathy}}, \bibinfo
  {author} {\bibfnamefont {G.~E.}\ \bibnamefont {Granroth}}, \bibinfo {author}
  {\bibfnamefont {T.~E.}\ \bibnamefont {Sherline}}, \bibinfo {author}
  {\bibfnamefont {P.~E.}\ \bibnamefont {Sokol}}, \bibinfo {author}
  {\bibfnamefont {J.}~\bibnamefont {Hu}},\ and\ \bibinfo {author}
  {\bibfnamefont {M.}~\bibnamefont {Boninsegni}},\ }\bibfield  {title}
  {\bibinfo {title} {Zero-point motion of liquid and solid hydrogen},\ }\href
  {https://doi.org/10.1103/PhysRevB.107.094511} {\bibfield  {journal} {\bibinfo
   {journal} {Phys. Rev. B}\ }\textbf {\bibinfo {volume} {107}},\ \bibinfo
  {pages} {094511} (\bibinfo {year} {2023})}\BibitemShut {NoStop}%
\bibitem [{\citenamefont {Myung}\ \emph {et~al.}(2022)\citenamefont {Myung},
  \citenamefont {Hirshberg},\ and\ \citenamefont {Parrinello}}]{parinello}%
  \BibitemOpen
  \bibfield  {author} {\bibinfo {author} {\bibfnamefont {C.~W.}\ \bibnamefont
  {Myung}}, \bibinfo {author} {\bibfnamefont {B.}~\bibnamefont {Hirshberg}},\
  and\ \bibinfo {author} {\bibfnamefont {M.}~\bibnamefont {Parrinello}},\
  }\bibfield  {title} {\bibinfo {title} {Prediction of a supersolid phase in
  high-pressure deuterium},\ }\href
  {https://doi.org/10.1103/PhysRevLett.128.045301} {\bibfield  {journal}
  {\bibinfo  {journal} {Phys. Rev. Lett.}\ }\textbf {\bibinfo {volume} {128}},\
  \bibinfo {pages} {045301} (\bibinfo {year} {2022})}\BibitemShut {NoStop}%
\end{thebibliography}%

\end{document}